\documentclass[prl, aps, twocolumn, floatfix, superscriptaddress, amsmath,  longbibliography]{revtex4-1}
\usepackage{graphicx}
\usepackage{graphics}
\usepackage{braket}
\usepackage{mathtools}
\usepackage[titletoc]{appendix}
\usepackage{appendix}
\usepackage{float}

\usepackage[dvipsnames]{xcolor}
\definecolor{darkblue}{rgb}{0.0, 0.0, 0.75}

\usepackage{color}
\usepackage[colorlinks=true,
            linkcolor=darkblue,
            urlcolor=darkblue,
            citecolor=darkblue]{hyperref}

\def \bk{{\bf k}}
\def \br{{\bf r}}
\def \bv{{\bf v}}

\def \bj{{\bf j}}

\def \mum{\mu \mathrm{m}}

\def \m2D{\mathrm{2D}}
\def \mB{\mathrm{B}}
\def \mms{\mathrm{ms}}
\def \mmms{\mathrm{mm/s}}

\def \mqc{\mathrm{qc}}

\def \mF{\text{F}}
\def \mp{\text{p}}

\def \mHz{\mathrm{Hz}}

\def \cD{\mathcal{D}}

\def \cH{\mathcal{H}}

\def \mmax{\mathrm{max}}

\def \bk{\mathbf{k} }
\def \br{\mathbf{r} }
\def \bq{\mathbf{q} }

\def \hn{\hat{n} }

\def \vt{ {\tilde{v} }}

\def \v0{ {\tilde{v}_0 }}

\def \bA{\textbf{A}}
\def \bB{\textbf{B}}
\def \bC{\textbf{C}}
\def \bD{\textbf{D}}
\def \bE{\textbf{E}}
\def \bF{\textbf{F}}

\def \cO{\mathcal{O} }

\def \2F{{_2}F}

\def \gt {\tilde{g} }

\begin{document}
\title{Collective modes and superfluidity of a two-dimensional ultracold Bose gas}
\author{Vijay Pal Singh and Ludwig Mathey}
\affiliation{Zentrum f\"ur Optische Quantentechnologien and Institut f\"ur Laserphysik, Universit\"at Hamburg, 22761 Hamburg, Germany}
\affiliation{The Hamburg Centre for Ultrafast Imaging, Luruper Chaussee 149, Hamburg 22761, Germany}
%
%
\date{\today}
%
%
\begin{abstract}
The collective modes of a quantum liquid shape and impact its properties profoundly, including its emergent phenomena such as superfluidity. Here we present how a two-dimensional Bose gas responds to a moving lattice potential. 
In particular we discuss how the induced heating rate depends on the interaction strength and the temperature. 
This study is motivated by the recent measurements of Sobirey {\it et al.} arXiv:2005.07607 (2020),  for which we provide a quantitative understanding. Going beyond the existing measurements, we demonstrate that this probing method allows to identify first and second sound in quantum liquids. We show that the two sound modes undergo hybridization as a function of interaction strength, which we propose to detect experimentally. 
This gives a novel insight into the two regimes of Bose gases, defined via the hierarchy of sounds modes. 
\end{abstract}

\maketitle
%
%

\section{Introduction}

The emergent phenomena of quantum liquids such as superfluidity and sound modes depend on a multitude of system features, such as interaction strength, dimensionality and temperature. These two classes of phenomena are linked in an intricate manner, as exemplified by the Landau criterion, which predicts dissipationless flow for a perturbation moving with a velocity below a critical velocity. This critical velocity in turn is associated with the creation of elementary excitations \cite{Kapitza1938, Misener1938, Allum1977, Pickett1986, McClintock1995}.
While study of superfluids was first motivated by the properties of helium-II, ultracold atoms have expanded the scope of the study of superfluidity by a wide range of trappable quantum liquids including superfluids having reduced dimensionality  and tunable interactions. 
Measurements of the critical velocity have been performed by perturbing ultracold atom clouds with a moving laser beam \cite{Ketterle1999, Dalibard2012, Weimer2015} or  lattice potential \cite{Ketterle2007, Chin2015}.

Superfluidity in two-dimensional (2D) systems is a particularly intriguing case, due to their critical properties, which differ from higher dimensional systems. 
Although 2D systems have no long-range order due to increased thermal fluctuations, they can become a superfluid via the Berezinksii-Kosterlitz-Thouless (BKT) mechanism \cite{Berezinski1972, Kosterlitz1973, Minnhagen1987}.  
The superfluid phase is a quasicondensate characterized by an algebraically decaying phase coherence. 
Ultracold atom systems provide unprecedented control and tunablity, allowing the study of superfluidity in 2D.
This led to the observation of pair condensation \cite{Ries2015}, phase coherence \cite{Hadzibabic, Murthy2015, Luick2020}, critical velocity \cite{Dalibard2012, Lennart}, and sound propagation  \cite{Dalibard2018, Markus2020}.
Two sound modes were recently detected in Ref. \cite{Hadzibabic2020}.
However, their hybridization and sublinear dissipation for small perturbation velocities, which we identify in this paper, have not been detected yet.

\begin{figure}[]
\includegraphics[width=1.0\linewidth]{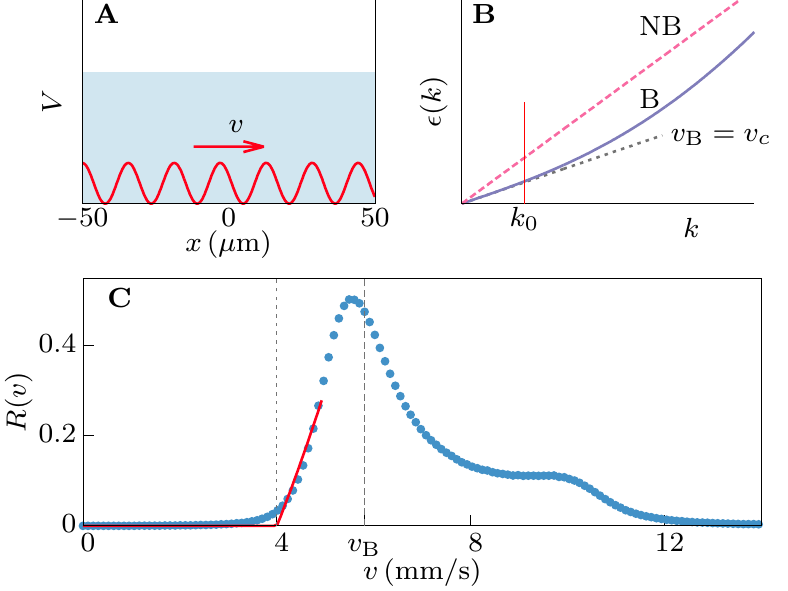}
\caption{\textbf{Superfluid response.}  
(\bA) A 2D superfluid in a box potential is probed by moving a lattice potential with lattice vector $k_0$ through it, at a velocity $v$. 
(\bB) The resulting heating of the superfluid derives from the excitation branches at $k_0$. 
For the weak-coupling regime, for which the velocity of the non-Bogoliubov (NB) mode is larger than the velocity $v_\mB$ of the Bogoliubov (B) mode, the Landau criterion predicts the critical velocity $v_c = v_\mB$, as indicated.
(\bC) Simulated heating rate $R(v)$ for $k_0/k_c = 0.6$, where $k_c = 2.2 \, \mum^{-1}$ is the wave vector above which the Bogoliubov dispersion approaches a quadratic momentum dependence. 
The fit (red curve) yields a sharp increase at $v_c= 4.0\, \mmms$,  see text.
The two maxima of the heating rate correspond to the two modes, where the first maximum is close to $v_\mB=5.8\, \mmms$.
 }
\label{Fig1}
\end{figure}

In this article, we use both classical-field dynamics and analytical estimates to investigate the induced heating rate as a function of the velocity of a moving lattice potential.  
The simulated heating rate shows two maxima corresponding to two sound modes, and sublinear dissipation at low velocities.
By changing the periodicity of the lattice potential we examine the heating rate at varying wave vector. 
We find phononic excitations for low wave vectors and free-particle excitations for high wave vectors.
This is in excellent agreement with Bogoliubov theory and the measurements of Ref. \cite{Lennart}. 
We determine a critical velocity from the sharp onset of dissipation and compare it with the measurements for various interactions \cite{Lennart}. 
Below the critical velocity, we find that the heating rate has a power-law dependence in the velocity and its magnitude increases with the temperature, which is supported by the analytic estimate of the quasicondensed phase. 
Finally, we determine the two mode velocities from the heating rate and identify their hybridization dependent on  interaction and temperature.

\section{Results}

\subsection{System and dynamical response}
We simulate bosonic clouds of $^{6}\mathrm{Li}_2$ molecules confined to 2D motion in a box potential. 
This geometry offers tunability of the effective interaction strength, and was used in Ref. \cite{Lennart}, and depicted in Fig. \ref{Fig1}A.
The system is perturbed by a lattice potential moving at a constant velocity $v$. 
The unperturbed system is described by the Hamiltonian 
\begin{equation} \label{eq_hamil}
\hat{H}_{0} = \int d \br \, \Big[\frac{\hbar^2}{2m} \nabla \hat{\psi}^\dagger({\bf r})  \cdot \nabla \hat{\psi}({\bf r}) 
 + \frac{g}{2} \hat{\psi}^\dagger({\bf r})\hat{\psi}^\dagger({\bf r})\hat{\psi}({\bf r})\hat{\psi}({\bf r})\Big].
\end{equation}
$\hat{\psi}$ ($\hat{\psi}^\dagger$) is the bosonic annihilation (creation) operator.
The 2D interaction parameter is $g = \tilde{g}\hbar^2/m$, where $\tilde{g}= \sqrt{8 \pi} a_s/\ell_z$ is the dimensionless interaction, 
$m$ the molecular mass, $a_s$ the 3D molecular scattering length, and $\ell_z$ the harmonic oscillator length in the transverse direction \cite{Turlapov2017}. 
We consider a rectangular system with dimensions $L_x \times L_y = 100 \times 100 \, \mum^2$ and a density  $n=1.2\, \mum^{-2}$, comparable to the experimental parameters \cite{Lennart}. 
We solve the dynamics using the classical-field method described in Refs. \cite{Singh2017, SinghJJ}, see also Methods.
According to this methodology, we replace the operators $\hat{\psi}$ in Eq. \ref{eq_hamil} and in the equations of motion by complex numbers $\psi$.
We sample the initial states from a grand-canonical ensemble with a chemical potential $\mu$ and a temperature $T$ via a classical Metropolis algorithm. 
To obtain the many-body dynamics of the system, we propagate the state using the classical equations of motion. 
We model the lattice perturbation via the additional term
\begin{align}\label{eq:Hd}
 \cH_\mp = \int d \br V(\br, t) n(\br),
\end{align}
where $n(\br)$ is the density at the location $\br = (x, y)$.  
The lattice potential $V(\br, t)$ is directed along the $x$ direction:  
\begin{align}\label{eq:pot}
V(x, t) = V_0 \cos^2[k_0(x+vt)/2].
\end{align}
$V_0$ is the strength, $k_0 = 2\pi/\lambda_l$ the wavevector, and $v$ the velocity, 
where we define $\lambda_l$ as the distance between two maxima of the potential.  
We move this potential for a fixed perturbation time $t_\mp$ of  $100\, \mms$. 
We calculate an ensemble average of the energy change $\Delta E = \langle H_0(t_\mp)\rangle- \langle H_0(0) \rangle$ using Eq. \ref{eq_hamil} and $\psi (\br, t)$. 
From this change of energy, we determine the heating rate $\Delta E/t_\mp$ for various sets of parameters $v$, $k_0$, $V_0$,  $\gt$,  and $T/T_0$. 
Throughout this paper, we will use the temperature $T_0 = 2\pi n \hbar^2 /(m k_\mB \cD_c)$, with the critical phase-space density $\cD_c= \ln(380/\tilde{g})$, as the temperature scale, see Refs. \cite{Prokofev2001, Prokofev2002}. 
This scale gives an estimate of the critical temperature $T_c$, which is the temperature of the BKT transition.
We define a dimensionless heating rate $R= \hbar \Delta E/(t_\mp V_0^2 N)$,  where $N$ is the number of  molecules. 
This heating rate and its velocity, temperature and interaction dependence are central quantities of this paper. 
As an example, in Fig. \ref{Fig1}C we show $R(v)$ for $\gt=1$ and $T/T_0=0.3$. 
We used $V_0/\mu = 0.05$ and  $k_0/k_c = 0.6$. 
$\mu = gn$ is the mean-field energy and $k_c \equiv \sqrt{2}/\xi$ is the wave vector above which the Bogoliubov dispersion approaches a quadratic momentum dependence,  with $\xi=\hbar/\sqrt{2m gn}$ being the healing length.  
As we depict in Fig. \ref{Fig1}C, the heating rate is small at low velocities. Below, we comment on the velocity dependence in this regime in more detail. 
The heating rate increases rapidly around a velocity which we refer to as the critical velocity of the condensate. 
As a simple estimate of this velocity, we fit the heating rate below the Bogoliubov velocity $v_\mB= \sqrt{gn/m}=5.8\, \mmms$ to the function $f(v)= A_0 \mmax[0, v^2 -v_c^2]$, with $A_0$ and $v_c$ as fitting parameters.
The fit gives $v_c=4.0\, \mmms$, as depicted in Fig. \ref{Fig1}C. 


  In addition to the sudden increase of the heating rate, captured by the critical velocity, the heating rate displays two maxima. These two maxima derive from the two excitation branches of the system. In the example shown in Fig. \ref{Fig1}C, the maximum at lower velocities is close to the Bogoliubov estimate $v_\mB$, as shown. We give further evidence for this interpretation below. We refer to this branch as the Bogoliubov (B) mode. Additionally, there is a second maximum at a higher velocity, which we refer to as the non-Bogoliubov (NB) mode. 
This is consistent with an excitation spectrum sketched in Fig. \ref{Fig1}B. 
Characterizing these two modes further is the second objective of this paper, in addition to the critical velocity.

 \begin{figure*}[]
\includegraphics[width=1.0\linewidth]{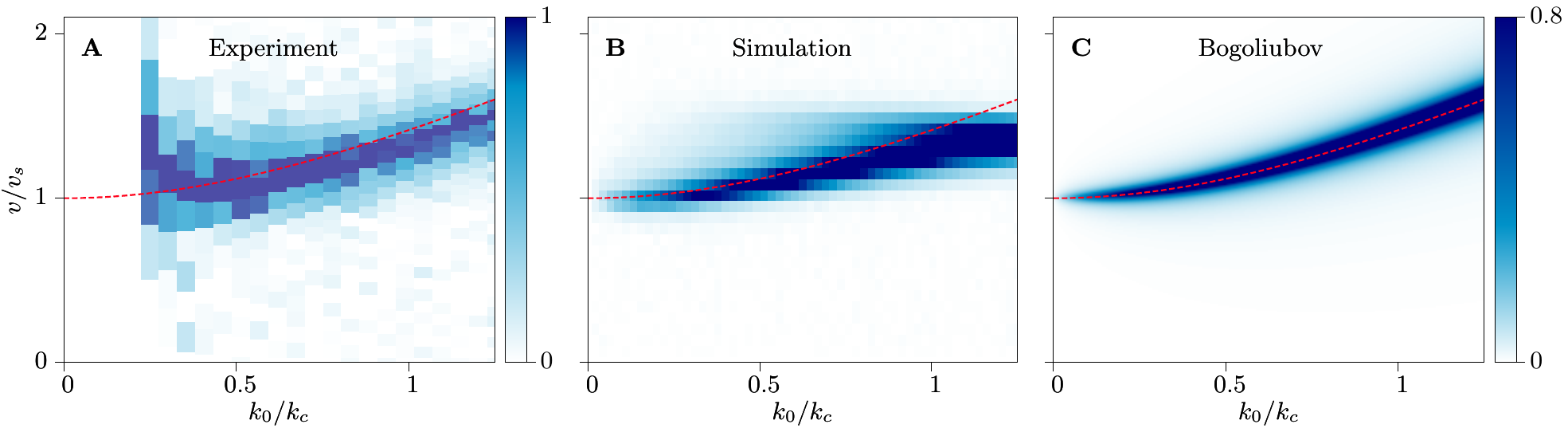}
\caption{\textbf{Heating rate dependence on the lattice vector.}
(\bA)  Measurements of the response $r(v)$ of Ref. \cite{Lennart}, determined from the heating of the condensate density and normalized individually for each column, as a function of $k_0/k_c$ and $v/v_s$.  
The sound velocity $v_s$ is determined from the propagation of a density wave \cite{SM1}. 
(\bB) Simulated heating rate $R(v)$ and (\bC) Bogoliubov estimate $R_\mB (v)$, for the same interaction and the same lattice parameters as in the experiment. 
The density $n = 1.2\, \mum^{-2}$, the interaction strength $\tilde{g}=1.6$, and the temperature $T/T_0=0.1$ are close to the experiments.
The red dashed line is the result $v_\mmax/v_B= \sqrt{1+(k_0/k_c)^2}$ of Eq. \ref{eq:BogM2}.  
The experimental result is primarily due to the Bogoliubov mode, reflected in the good agreement with the simulation and the analytical estimate.
   }
\label{Fig:phonon}
\end{figure*}

\begin{figure}[]
\includegraphics[width=1.0\linewidth]{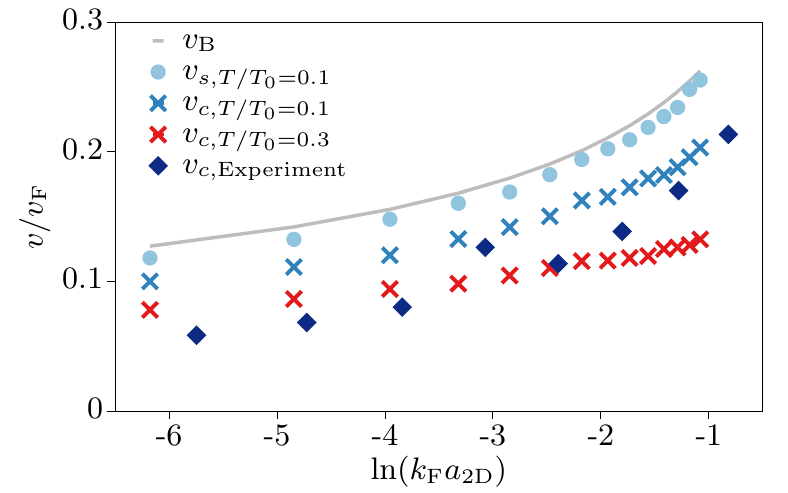}
\caption{\textbf{Critical velocity.} 
Measurements of the critical velocity (diamonds) of Ref. \cite{Lennart} are compared to the simulations of $T/T_0=0.1$ (blue crosses) and $0.3$ (red crosses) for various interactions $\ln(k_\mF a_\m2D)$. 
$k_\mF$ is the Fermi wave vector, $a_\m2D$ is the 2D scattering length, and $v_\mF$ is the Fermi velocity.
We also show the propagation velocity of a density wave (circles) for $T/T_0=0.1$ and the Bogoliubov velocity $v_\mB$ (continuous line).
}
\label{Fig:vc}
\end{figure}

\subsection{Analytical estimates} 
  We present two analytical estimates for properties of the heating rate. The first estimate uses Bogoliubov theory, and the second estimate uses the quasicondensate properties of 2D quantum gases. 
We derive the heating rate perturbatively at second order in the probing term. 
The Bogoliubov estimate of the heating rate is  \cite{SM1}
\begin{equation} \label{eq:BogM}
\frac{d E}{dt} = \frac{2\pi}{\hbar} \sum_{\bf k} \omega_k (u_k + v_k)^2 N_0 |V_\bk|^2  \delta(\omega_k - {\bf vk}).
\end{equation}
$\hbar \omega_k = \sqrt{\epsilon_k (\epsilon_k  + 2 m v_\mB^2)}$ is the Bogoliubov dispersion, $u_k$ and $v_k$ are the Bogoliubov parameters, 
with $(u_\bk + v_\bk)^2 = \epsilon_k/\hbar \omega_k$. $\epsilon_k= \hbar^2 k^2/(2m)$ is the free-particle dispersion. 
$V_k=V_0 \delta_{k_y} ( \delta_{k_x - k_0}  + \delta_{k_x +  k_0}  )/4$ is the time independent part of the lattice potential in momentum space, and $N_0$ is the number of condensed atoms. 
In dimensionless form $R \equiv \hbar (dE/dt)/(N_0 V_0^2)$ and after simplifying Eq. \ref{eq:BogM} we obtain 
\begin{align}\label{eq:BogM2}
R_\mB = \frac{\pi}{16}  \frac{\hbar k_0}{m}  \delta[ \sqrt{v_0^2/4 + v_\mB^2} - v],
\end{align}
where $v_0$ is $v_0 = \hbar k_0/m$.  The normalized heating rate $R_\mB$ has a maximum at $v_\mmax= (v_0^2/4 + v_\mB^2)^{1/2}$. 
The location of the maximum moves to higher velocities with increasing $k_0$ or $v_0$.
We show $v_\mmax$ as a red line in Fig. \ref{Fig:phonon}, which captures the trend of the measurement and the simulation.
We include the thermal damping of the phonon modes by considering a Landau-type damping $\Gamma_k = v_\Gamma k$, where $v_\Gamma$ is the damping velocity. 
This results in  \cite{SM1}
\begin{align}\label{eq:BogTemp}
R_\mB = \frac{1}{16} \frac{v_0 v_\Gamma}{ ( \sqrt{v_0^2/4 + v_\mB^2} - v)^2 + v_\Gamma^2 }. 
\end{align} 
We show this estimate in Fig. \ref{Fig:phonon}C.

  To describe the heating rate at low velocities we relate the heating rate to the momentum distribution $n_\bk$, and arrive at the expression \cite{SM1}
\begin{equation}\label{eq:full}
\frac{d E}{dt}  = \frac{2\pi}{\hbar^2} \sum_{\bk \bq} (E_{\bk + \bq} - E_{\bq} ) n_\bq  |V_\bk|^2  \delta( {\bf vk} + \omega_\bq - \omega_{\bk + \bq}   ),
\end{equation}
where $E_\bk = \hbar v_\mB |\bk|$ is the phononic dispersion at long wavelengths.  
The momentum density $n_\bk$ follows a power-law dependence in the wave vector $\bk$ as
\begin{align}\label{eq:nk}
n_\bk = n \frac{\pi \tau}{2}  r_0^{\tau/4} |\bk|^{\tau/4 - 2 },
\end{align}
where $r_0$ is the short-range cutoff of the order of $\xi$ and $\tau$ is the algebraic scaling exponent.
$\tau$ increases monotonously from $0$ to $1$ as the temperature is increased from $0$ to $T_c$. 
In dimensionless form and for $v < v_\mB$, we obtain the heating rate of a quasicondensate at low temperatures \cite{SM1}
\begin{align}\label{eq:Rqc}
R_\mqc = \frac{\pi}{32} \frac{\tau  v^2}{v_\mB^2 - v^2},
\end{align}
which scales as $R_\mqc \propto \tau v^2 $ for low $v$ and yields a rapid increase for $v$ close to $v_\mB$. It vanishes, as $\tau$ approaches $0$.

\begin{figure}[t]
\includegraphics[width=1.0\linewidth]{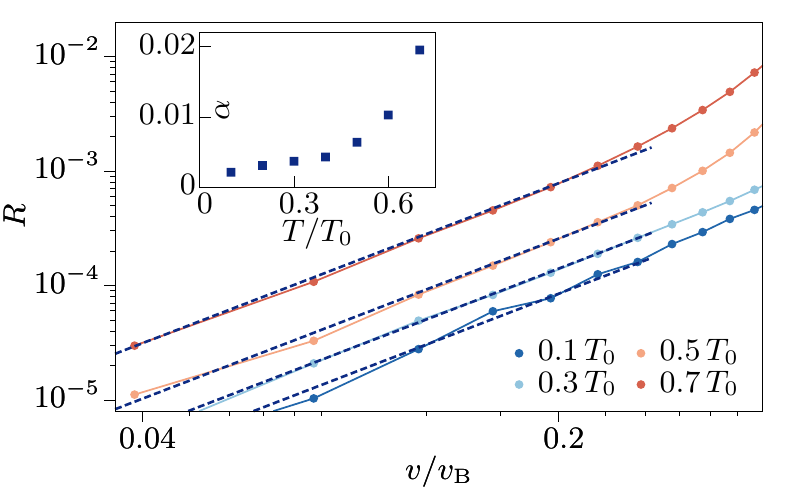}
\caption{\textbf{Low-velocity dependence of the heating rate.}  
Simulated heating rate $R(v)$ on a log-log scale for a weak perturbation and various $T/T_0$.  
The dashed lines are the algebraic fits with the function $f( \tilde{v} )= \alpha \tilde{v}^2$, where $\tilde{v}=v/v_\mB$ is the scaled velocity. 
 The fit parameter $\alpha (T)$ is shown in the inset. 
 }
\label{Fig:qc}
\end{figure}

\begin{figure}[t]
\includegraphics[width=1.0\linewidth]{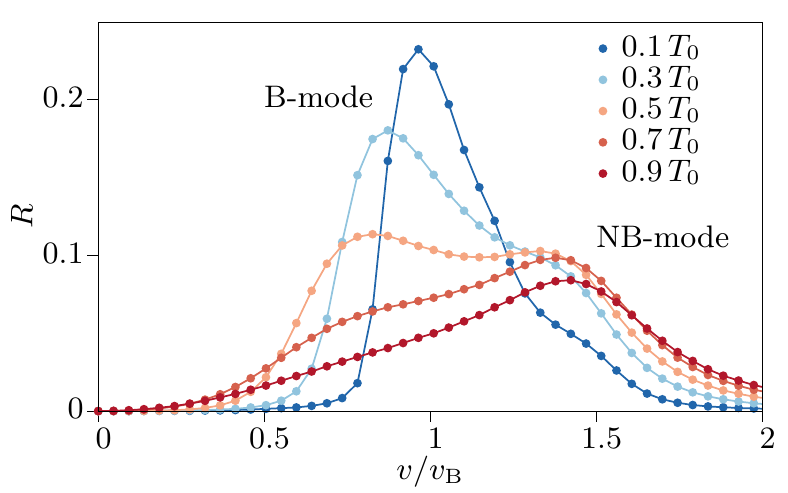}
\caption{\textbf{Two sound modes.}  
$R(v)$ for $\gt=1.4$,  $k_0/k_c = 0.4$ and various $T/T_0$. 
The transition temperature is $T_c/T_0 = 0.86$, which is determined from a zero critical velocity  \cite{SM1}.
  }
\label{Fig:temp}
\end{figure}

 \begin{figure*}[]
\includegraphics[width=1.0\linewidth]{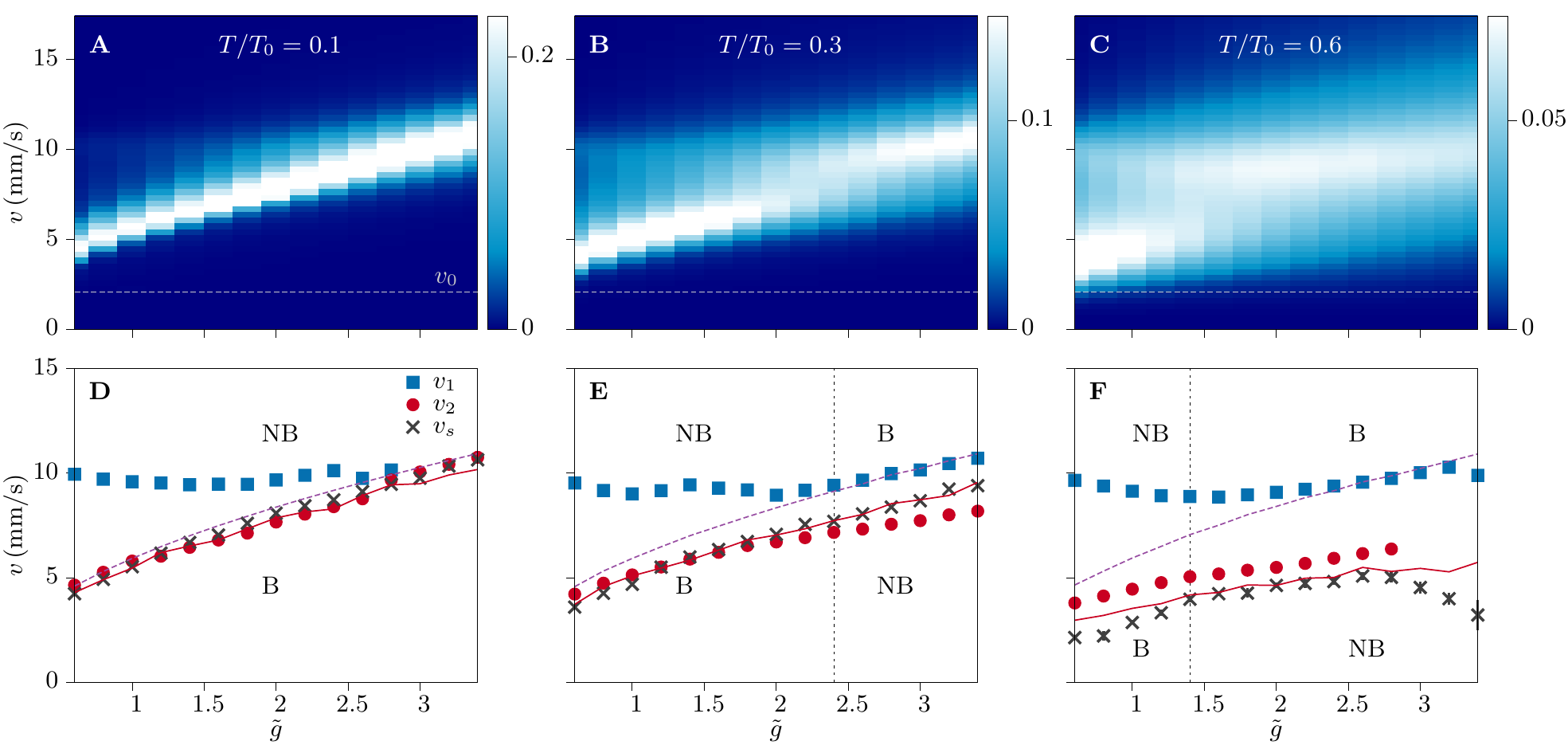}
\caption{\textbf{Hybridization of the two sound modes.}  (\bA,\bB,\bC) $R(v)$ as a function of $\gt$ for $T/T_0=0.1$, $0.3$, and $0.6$.  
The horizontal dashed line denotes $v_0=2\, \mmms$ corresponding to $k_0 = 1.1\, \mum^{-1}$. 
(\bD,\bE,\bF) The two mode velocities $v_{1/2}$ determined from the heating rate are compared with the propagation velocity of density waves (crosses),  and the Bogoliubov estimates employing the total density (dashed line) and the numerical superfluid density (continuous line). 
B denotes the Bogoliubov sound mode and NB the non-Bogoliubov mode. 
The vertical dashed line indicates the hybridization point, see text. 
 }
\label{Fig:modes}
\end{figure*}

\subsection{Comparison} 
In this section, we compare the analytical estimates with the simulation results and the experimental results. 
In Fig. \ref{Fig:phonon}A we show the measurements of Ref. \cite{Lennart}.
The measurement was performed at $\ln(k_\mF a_\m2D) = -2.8$, where $k_\mF=\sqrt{4\pi n}$ is the Fermi wave vector and $a_\m2D$ is the 2D scattering length.
The maximum of this measured response is close to the phonon velocity for small $k_0/k_c$ and shifts to larger velocities with increasing $k_0/k_c$.
We perform a simulation for the same system parameters, which results in the heating rate shown in Fig. \ref{Fig:phonon}B. 
The heating rate displays the same overall dependence on the lattice wave vector $k_0$. In particular, for vanishing $k_0$ the maximum of the heating rate approaches the sound velocity.
Furthermore, the simulation recovers the measurements for intermediate and high wave vectors.  
We note that for $k_0/k_c >1 $ the lattice wave vector $k_0$ approaches the maximum momentum set by the system discretization length, 
so that the simulation becomes quantitatively unreliable.
In Fig. \ref{Fig:phonon}C we show the Bogoliubov estimate $R_\mB$ of Eq. \ref{eq:BogTemp}. 
We set the value of $v_\Gamma/v_\mB=0.03$, that we have obtained numerically by determining the damping of a density wave \cite{SM1}.
The magnitude of the maximum of $R_\mB$ increases with increasing $k_0$ as in the simulation in Fig. \ref{Fig:phonon}B, and provides a good estimate for both the measurement and the simulation. 
We note that the simulation also displays a faint second branch above the Bogoliubov branch. As we elaborate below, this is due to the non-Bogoliubov mode. This feature becomes more pronounced and realistically detectable for higher temperatures and for an interaction strength near the hybridization interaction, as we discuss below.

      Next, we determine the critical velocity for the same value $k_0/k_\mF=0.3$ and for varying interactions as in the experiment. 
In Fig. \ref{Fig:vc} we compare the simulation results of the critical velocity $v_c$ for $T/T_0 =0.1$ and $0.3$ with the measurements of Ref. \cite{Lennart}.  
The measurement agrees with the simulation of $T/T_0=0.3$ for the interaction strengths  $\ln(k_\mF a_\m2D) \lesssim -1.5$, except for the data point  at  $\ln(k_\mF a_\m2D) = -3.2$.  For strong interactions $\ln(k_\mF a_\m2D) \gtrsim -1.5$, the measurements are closer to the simulations of  $T/T_0=0.1$.  
This might suggest that the experimental results were obtained at temperatures that varied with varying interaction strengths.
We also show the simulation results of the sound velocity $v_s$ for $T/T_0=0.1$,  which is determined from the propagation of a density wave \cite{SM1}. 
For all interactions, this sound velocity is slightly below the Bogoliubov velocity $v_\mB$ and above the results of $v_c$. 
The difference between $v_s$ and $v_c$  is higher for strong interaction and high temperature, which is due to the broadening of the heating rate.

   Finally, we examine the low-velocity behavior of the heating rate at $v/v_\mB \ll 1$. 
In Fig. \ref{Fig:qc} we show the simulated heating rate $R(v)$ for $k_0/k_c =0.4$ and various $T/T_0$. 
$R(v)$ shows a power-law dependence at low velocities, visible as a linear dependence on $v$ on a log-log scale. 
More specifically, we observe a quadratic dependence on $v$, as supported by fitting with the function $f(\tilde{v})= \alpha \tilde{v}^2$, 
where $\alpha$ is the fitting parameter. 
$\tilde{v}=v/v_\mB$ is the scaled velocity. 
We refer to this power-law dependence as sublinear dissipation, because the power-law dependence is sublinear. 
We note that above the critical temperature, where the bosons form a thermal cloud, the dissipation is linear in $v$ \cite{Singh2016}. 
The quantity $\alpha(T)$ increases with the temperature, as we show in the inset of Fig. \ref{Fig:qc}. 
Both the power-law dependence and $\alpha(T)$ are consistent with the estimate $R_\mqc$ in Eq. \ref{eq:Rqc}, 
where $\alpha(T)$ is related to the BKT exponent $\tau$.
Since the magnitude of the heating rate is small, it would be demanding to extract $\tau$ experimentally.
However, in principle this result relates the low-velocity heating rate to the quasicondensate scaling exponent.

\subsection{Two sound modes}
    In this section we characterize the two sound modes that we observe in the heating rate.
In Fig. \ref{Fig:temp} we show the simulation results of $R(v)$ for $\gt=1.4$, $k_0/k_c=0.4$, and various $T/T_0$.       
$R(v)$ displays two maxima. The first maximum corresponds to the B mode and the second maximum to the NB mode. 
At low temperatures the B mode has a higher magnitude than the NB mode. This changes as the temperature is increased. 
For high temperatures the NB mode has a higher magnitude than the B mode. 
This shift of the magnitude is consistent with the shift of spectral weights of the two modes in the dynamic structure factor \cite{SinghSS}. 
Thus, the heating rate provides direct access to the relative amplitude of two modes \cite{Hoffmann}. 
The first maximum disappears and turns into a broad background of the diffusive sound mode above the critical temperature $T_c/T_0= 0.86$, 
where the value of $T_c$ is identified by the critical velocity going to zero \cite{SM1}.

      Next, we analyze the interaction dependence by varying $\gt$ in the range $0.6-3.4$ and using the lattice wave vector fixed at $k_0 = 1.1\, \mum^{-1}$.
In Figs. \ref{Fig:modes}(A-C) we show $R(v)$ as a function of $\gt$ for $T/T_0 = 0.1$, $0.3$, and $0.6$.  
$R(v)$ displays a hybridization of the two sound modes as a function of $\gt$. 
This hybridization is more visible at intermediate and higher temperatures, due to the larger weight of the non-Bogoliubov mode, which we observed in Fig. \ref{Fig:temp}.
For low temperatures, as in Fig. \ref{Fig:modes}A, this hybridization occurs at a high value of $\gt$, which shifts to low values of $\gt$ for high temperatures in Figs. \ref{Fig:modes}B and \ref{Fig:modes}C. 
Similarly, the magnitude of the velocity difference at the avoided crossing increases with increasing temperature.
To study this hybridization we determine the two mode velocities by fitting $R(v)$ with a single and a bi-Lorentzian function.
In Figs. \ref{Fig:modes}(D-F) we present the interaction dependence of the two velocities $v_{1/2}$ obtained from the fit.
We indicate the hybridization point at the interaction at which the magnitude of the heating rate of the high-velocity ($v_1$) mode exceeds the magnitude of the heating rate of the low-velocity ($v_2$) mode by a vertical dashed line in Figs. \ref{Fig:modes}E and  \ref{Fig:modes}F. 
We compare $v_{1/2}$ with the velocity $v_s$ obtained from the propagation of a density wave \cite{SM1}.
We also show the Bogoliubov velocities $v_\mB$ and $v_{\mB, T}= \sqrt{g n_s(T)/m}$ determined using either the total density or the numerically obtained superfluid density $n_s(T)$ \cite{SM1}. 
The lower velocity $v_2$ agrees with $v_s$ and $v_{\mB,T}$ for all interactions and all temperatures. 
The upper velocity $v_1$ agrees with $v_\mB$ above the hybridization point only. 
This suggests that for weak interaction and low temperature the system is in the non-hydrodynamic regime, where the lower velocity is described the Bogoliubov (B) approach and the upper mode is the non-Bogoliubov (NB) mode, as we pointed out in Refs. \cite{Ilias, SinghSS}. 
This scenario changes for strong interaction and high temperatures, 
where the system enters the hydrodynamic regime and a hydrodynamic two-fluid model describes the two sound modes. 
Here, the upper mode propagates with the total density and the lower mode with the superfluid density \cite{Pethick2008, Pitaevskii}.

\section{Discussion} 

  We have determined and discussed the heating rate of a superfluid 2D Bose cloud by perturbing it with a moving lattice potential, 
using classical-field simulations and analytical estimates. 
This study is primarily motivated by the experiment reported in Ref. \cite{Lennart}. 
Indeed, we find, firstly, that the signature of the Bogoliubov mode in the heating rate is well-reproduced in our study, for which we present a numerical result as well as an analytical estimate. Secondly, we show that  the critical velocity that emerges in this type of stirring experiment is consistent with the experimental findings of Ref. \cite{Lennart}.

However, our study also suggests to broaden the scope of this type of stirring experiment. The results that we report here, elucidate the general behavior of sound excitations in 2D Bose gases, in particular they give access to the two-mode structure of the excitation spectrum, and their interaction and temperature dependence. We show that the heating rate has two maxima, as a function of its velocity and for fixed lattice wave length, corresponding to the two sound modes of the fluid. The relative weight of these modes changes significantly as a function of temperature, and reverts its hierarchy.  Furthermore, we show that the two modes undergo hybridization as a function of the interaction strengths. At interaction strengths below the hybridization strength, the non-Bogoliubov mode has a higher velocity than the Bogoliubov mode, whereas for interaction strengths above the hybridization strengths the hierarchy is reversed, which is consistent with the scenario described in \cite{SinghSS}. 
This provides a new insight into the collective modes of Bose gases, which we put forth here, and which we propose to be tested experimentally.

\section{Methods} 
To perform numerical simulations we discretize space using a lattice of size $N_x \times N_y = 200 \times 200$ and a discretization length $l= 0.5\, \mum$. 
We note that $l$ is chosen to be smaller than or comparable to the healing length $\xi=\hbar/\sqrt{2mgn}$ and the de Broglie wavelength  $\lambda= \sqrt{2\pi \hbar^2/(m k_\mB T)}$, see Ref. \cite{Castin2003}.
This results in the cloud size $L_x \times L_y=100 \times 100 \, \mum^{2}$.
For the system parameters we use the density $n=1.2\, \mum^{-2}$, and various interactions and temperatures.
The initial ensemble consists of  $100-1000$ states according to the temperature and the interaction. 
We propagate the state using the equations of motion. 
For perturbing the cloud, we linearly turn on the lattice potential over $100\, \mms$ and then move the potential at a velocity $v$ and a lattice wave vector $k_0$ for a fixed perturbation time $t_\mp$ of  $100\, \mms$.

\bibliography{References}

\section{Acknowledgements} 
We thank Lennart Sobirey, Thomas Lompe, and Henning Moritz for insightful discussions and providing us with the experimental data.  
This work was supported by the DFG in the framework of SFB 925 and the excellence clusters `The Hamburg Centre for Ultrafast Imaging’- EXC 1074 - project ID 194651731 and `Advanced Imaging of Matter’ - EXC 2056 - project ID 390715994.

\section{Supplementary materials} 

Supplementary Text \\

Fig. S1-S4 \\

References (17,  24, 28-29, 35)

\newpage
\widetext
\begin{center}
\textbf{\large Supplemental Materials: Collective modes and superfluidity of a two-dimensional ultracold Bose gas}
\end{center}
\setcounter{equation}{0}
\setcounter{figure}{0}
\setcounter{table}{0}
\makeatletter
\renewcommand{\theequation}{S\arabic{equation}}
\renewcommand{\thefigure}{S\arabic{figure}}
\renewcommand{\bibnumfmt}[1]{[S#1]}

\def \la{{\langle}}
\def \ra{{\rangle}}

\def \hH{\hat{H} }
\def \hb{\hat{b} }

\section{Analytic heating rate}\label{sec:rate}

We determine the heating rate peturbatively by considering a weak perturbation term. To second order, the heating rate is given by \cite{Singh2016}
\begin{align}\label{seq:second}
\frac{d \braket{ \hH_0 (t)} }{dt} = - \frac{1}{\hbar^2} \int_0^t d t_1  \langle \bigl[\hH_{s,I}(t_1), & [\hH_{s,I} (t), \hH_0] \bigr] \rangle.
\end{align}
$\hH_0$ is the unperturbed Hamiltonian. $\hH_{s, I}$ is the perturbation term in the interaction picture. 
In momentum space the perturbation term is described as  
\begin{align}\label{seq:term}
\hH_s = \sum_{\bk} V_\bk(t) \hn_\bk,
\end{align}
where $V_\bk(t)$ is the Fourier transform of the potential $V(\br, t)$ and $\hn_\bk = \sum_{\bq} a_{\bq}^\dagger a_{\bk +\bq}$  is the Fourier transform of the density operator $\hat{n}(\br)$.  $a_{\bk}$ ($a_{\bk}^\dagger$) is the bosonic annihilation (creation) operator. 
$V(\br, t)$ is the lattice potential directed along the $x$ direction: $V(x, t) = V_0 \cos^2[k_0(x+vt)/2]$, where $V_0$ is the strength, $v$ the velocity, and $k_0$ the wave vector. We Fourier transform $V(x, t)$ and obtain  
\begin{align}\label{seq:Vk}
V_{\bk}(t) = \frac{V_0}{4}\delta_{k_y} ( \delta_{k_x - k_0}  + \delta_{k_x + k_0}  ) \exp(-i k_x v t).
\end{align}
We transform Eq. \ref{seq:term} to the interaction picture via $\hH_{s, I} (t)= \exp( i \hH_0t )  \hH_s(t)  \exp(- i \hH_0 t) $.

\subsection{Bogoliubov heating rate}  
Here we derive the heating rate for a condensate at zero temperature. We use the Bogoliubov approximation and obtain the diagonalized Hamiltonian 
\begin{align}\label{seq:Hbog}
\cH_0 = \sum_\bk \hbar \omega_k \hb_\bk^\dagger \hb_\bk,
\end{align}
where $ \hb_\bk$ and $\hb_\bk^\dagger$ are the Bogoliubov operators, and $\hbar \omega_k = \sqrt{\epsilon_k (\epsilon_k + 2 m v_\mB^2)}$ is the Bogoliubov dispersion. $\epsilon_k= \hbar^2 k^2/(2m)$ is the free-particle dispersion and  $v_\mB$ is the Bogoliubov velocity. 
We expand the momentum occupation around the condensate mode as $\hn_\bk =   \sqrt{N_0} (u_\bk + v_\bk) (\hat{b}_{-\bk}^\dagger  + \hat{b}_{\bk} )$,  where $N_0$ is the number of condensed atoms, and $u_\bk$ and $v_\bk$ are the Bogoliubov parameters, with $(u_\bk + v_\bk)^2 = \epsilon_k/\hbar \omega_k$.  This results in $\cH_s (t) = \sum_{\bk} V_\bk(t) \sqrt{N_0} (u_\bk + v_\bk) (\hat{b}_{-\bk}^\dagger  + \hat{b}_{\bk} ) $. For the interaction picture we use $ \hat{b}_{\bk} \rightarrow \hat{b}_{\bk} \exp(- i \omega_k t) $ and  $ \hat{b}_{\bk}^\dagger \rightarrow \hat{b}_{\bk}^\dagger \exp( i \omega_k t) $, which yields
\begin{align}\label{seq:term2}
\cH_{s, I} (t) =   \sum_{\bk} V_\bk(t) \sqrt{N_0} (u_\bk + v_\bk) (\hat{b}_{-\bk}^\dagger e^{i \omega_k t} + \hat{b}_{\bk} e^{-i \omega_k t}  ).
\end{align}
Using Eqs. \ref{seq:Hbog} and \ref{seq:term2} we solve the commutator in Eq. \ref{seq:second} and arrive at the heating rate \cite{Singh2016}
\begin{equation} \label{seq:Bog}
\frac{d E}{dt} = \frac{2\pi}{\hbar} \sum_{\bf k} \omega_k (u_k + v_k)^2 N_0 |V_\bk|^2  \delta(\omega_k - \bv  \bk ).
\end{equation}
$V_\bk$ is the time-independent part of the lattice potential in Eq. \ref{seq:Vk}. Using the expression of $|V_k|^2$ and $\omega_k(u_k+ v_k)^2= \hbar k^2/(2m)$, we obtain 
 \begin{equation} \label{seq:Bog2}
\frac{d E}{dt} = \frac{\pi}{16}  \frac{N_0 V_0^2}{m}  k_0^2 \delta[ \omega_{k_0, 0} - v k_0 ]. 
\end{equation}
The delta term in the heating rate gives an onset of dissipation at the velocity $v= \omega_{k_0, 0}/k_0$, where $\omega_{k_0, 0} = k_0 \sqrt{v_0^2/4 + v_\mB^2}$ and $v_0=\hbar k_0/m$. In dimensionless form $R \equiv \hbar (dE/dt)/(N_0 V_0^2)$ we have
\begin{align}\label{seq:Bog3}
R_\mB = \frac{\pi}{16}  \frac{\hbar k_0}{m}  \delta[ \sqrt{v_0^2/4 + v_\mB^2} - v].
\end{align}
We extend this result to nonzero temperatures by including the thermal damping of the phonon modes. 
We consider a Landau-type damping $\Gamma_k= v_\Gamma k$, where $v_\Gamma$ is the damping velocity. 
We replace the delta distribution by a Lorentzian distribution, i.e $\pi \delta(x) = \lim_{ \epsilon \rightarrow 0} \epsilon /(x^2 + \epsilon^2)$. 
This results in 
\begin{align}\label{seq:BogTemp}
R_\mB = \frac{1}{16} \frac{v_0 v_\Gamma}{ ( \sqrt{v_0^2/4 + v_\mB^2} - v)^2 + v_\Gamma^2 }. 
\end{align} 
\subsection{Quasicondensate heating rate}  
\begin{figure}[]
\includegraphics[width=1.0\linewidth]{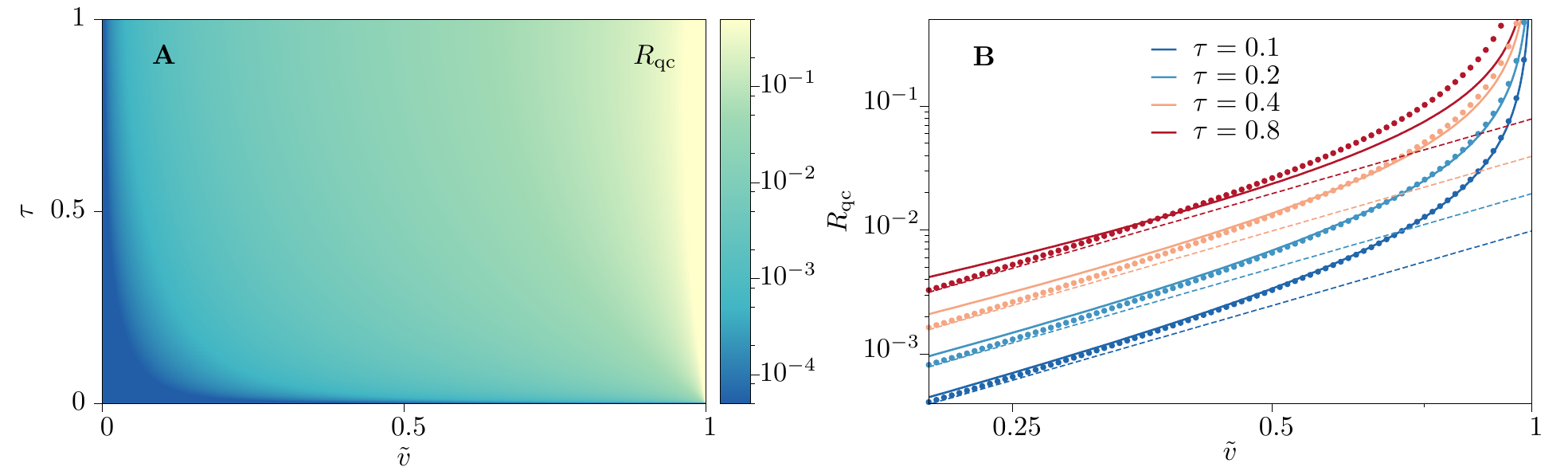}
\caption{\textbf{Heating rate for a quasicondensate.}  
(\bA) The estimate $R_\mqc$ of Eq. \ref{seq:SFfull} as a function of the scaled velocity $\vt=v/v_\mB$ and the algebraic scaling exponent $\tau$. The magnitude of the heating rate is displayed on a log scale.  
(\bB) The estimates of Eq. \ref{seq:SFfull} (continuous line),  Eq. \ref{seq:SFD} (dots), and  $R_\mqc= \pi  \tau \vt^2/32$ (dashed line) are plotted on a log-log scale, for $\tau=0.1$, $0.2$, $0.4$, and $0.8$.     }
\label{FigS:qc_rate}
\end{figure}

To derive the heating rate for a quasicondensate, we consider a Hamiltonian of the form
\begin{align} \label{seq:Hamil}
H_0 = \sum_{\bk} E_\bk  a_{\bk}^\dagger a_{\bk},
\end{align}
where $E_\bk \equiv \hbar \omega_\bk$ is the excitation spectrum. For the perturbation term, we transform Eq. \ref{seq:term} to the interaction picture by using $a_\bk \rightarrow \exp(-i \omega_k t) a_\bk$. This results in
\begin{align} \label{seq:pert2}
H_{s,I} = \sum_{\bk \bq} V_\bk(t) e^{i(\omega_\bq - \omega_{\bk+\bq})t}a_{\bq}^\dagger a_{\bk +\bq}.
\end{align}
We now use Eqs. \ref{seq:Hamil} and \ref{seq:pert2} to calculate the commutator $[H_{s,I} (t), H_0]$, which gives
\begin{align} \label{seq:comm_1}
[H_{s,I} (t), H_0] = \sum_{\bk \bq} (E_{\bk+\bq} - E_\bq)V_\bk(t) e^{i(\omega_\bq - \omega_{\bk+\bq})t} \times a_{\bq}^\dagger a_{\bk +\bq}.
\end{align}
Using Eqs. \ref{seq:pert2} and \ref{seq:comm_1} we calculate the commutator $\bigl[H_{s,I}(t_1),  [H_{s,I} (t), H_0] \bigr]$ and obtain
\begin{align} \label{seq:comm_2}
\bigl[H_{s,I}(t_1), & [H_{s,I} (t), H_0] \bigr] = - 2\sum_{\bk \bq} (E_{\bk+\bq} -E_\bq) |V_\bk|^2  \cos \bigl[(\bv \bk + \omega_\bq - \omega_{\bk+\bq})(t_1-t) \bigr]  a_{\bq}^\dagger a_{\bq}.
\end{align}
Integrating Eq. \ref{seq:comm_2} over time $t_1$ yields the heating rate 
\begin{align} \label{eq_non_rate_1}
\frac{dE}{dt} = \frac{2}{\hbar^2} \sum_{\bk \bq} (E_{\bk+\bq} -E_\bq) |V_\bk|^2 \frac{ \sin[(\bv \bk + \omega_\bq - \omega_{\bk+\bq}) t ]} {(\bv \bk + \omega_\bq - \omega_{\bk+\bq}) }  \la a_{\bq}^\dagger a_{\bq} \ra,
\end{align}
which at long times approaches
\begin{align}\label{seq:full}
\frac{d E}{dt}  = \frac{2\pi}{\hbar^2} \sum_{\bk \bq} (E_{\bk + \bq} - E_{\bq} ) n_\bq  |V_\bk|^2  \delta( {\bf vk} + \omega_\bq - \omega_{\bk + \bq}   ).
\end{align}
This result relates the heating rate to the momentum distribution $n_\bk$. 
We consider the phononic dispersion at long wave lengths, i.e., $E_\bk= \hbar v_\mB |\bk|$. 
The momentum distribution is given by
\begin{align}\label{seq:nk}
n_\bk = n \frac{\pi \tau}{2}  r_0^{\tau/4} |\bk|^{\tau/4 - 2 },
\end{align}
where $n$ is the real-space density, $r_0$ is the short-range cutoff of the order of $\xi$, and $\tau$ is the algebraic scaling exponent.  
We simplify Eq. \ref{seq:full} and obtain
\begin{align}\label{seq:SFfull}
\frac{d E}{dt}  = \frac{N V_0^2 }{128 \hbar}    (k_0 r_0/2)^{\tau/4}  \tau \vt (1- \vt^2)^{\tau/4 - 1 } I(\phi),
\end{align}
with 
\begin{equation}
I(\phi)  =  \int_0^{2\pi} d \phi  (1+\vt^2 -2\vt \cos  \phi )   (\vt- \cos \phi )^{-(1+ \tau/4)}.
\end{equation}
$\vt= v/v_\mB$ is the scaled velocity. $I(\phi) $ can be solved, giving
\begin{align}
&I(\phi)  =  \frac{\pi}{2} \Bigl[ 3(\vt-1)^{1-x} (\vt+1) \2F_1 \Bigl( - \frac{1}{2}, x, 2, - \frac{2}{\vt-1}  \Bigr) 
 + 3(\vt+1)^{-x} (\vt^2-1) \2F_1 \Bigl( - \frac{1}{2}, x, 2,  \frac{2}{\vt+1}  \Bigr) \nonumber \\
& - (\vt-1)^{-x} (\vt^2 +  4 \vt x+ 4x -5)  \2F_1  \Bigl(  \frac{1}{2}, x, 2, - \frac{2}{\vt-1}  \Bigr) 
 - (\vt+1)^{-x} (\vt^2 -4\vt x + 4x -5 )  \2F_1  \Bigl(  \frac{1}{2}, x, 2, \frac{2}{\vt+ 1}  \Bigr)  \Bigr],
\end{align}
where $x= (1+ \tau/4)$ and $\2F_1(a, b, c, d)$ are the hypergeometric functions. 
We expand $I(\phi) $ as $I(\phi) =  4 \pi \vt  + \cO(\tau)$. 
The dimensionless heating rate $R=\hbar(dE/dt)/(NV_0^2)$ is 
\begin{equation} \label{seq:SFD}
R_\mqc  =  \frac{\pi}{32}  \frac{ \tau \vt^2}{  1 - \vt^2}+  \cO(\tau).
\end{equation}
This low-temperature estimate scales linearly in $\tau$, while it vanishes for $\tau$ approaching zero. 
At this order the dependence of $k_0$ and $r_0$ drops out. 
For low velocity the heating rate scales quadratically in $v$ as $R_\mqc = \pi \tau \vt^2/32$.
In Fig. \ref{FigS:qc_rate}(a) we show the result of Eq. \ref{seq:SFfull} as a function of $\vt$ and $\tau$ for $k_0 r_0=2$.
The dissipation is nonzero below the Bogoliubov velocity and increases with increasing both $\vt$ and $\tau$.
This nonzero dissipation at low velocities is peculiar to 2D superfluids.
In Fig. \ref{FigS:qc_rate}(b) we compare the results of Eqs. \ref{seq:SFfull} and \ref{seq:SFD}, 
which show good agreement for all $\vt$ and all $\tau$.
The low-velocity sublinear behavior is also captured well by the estimate $R_\mqc = \pi \tau \vt^2/32$.

 \section{Influence of temperature and lattice strength on the critical velocity}
\begin{figure}[]
\includegraphics[width=1.0\linewidth]{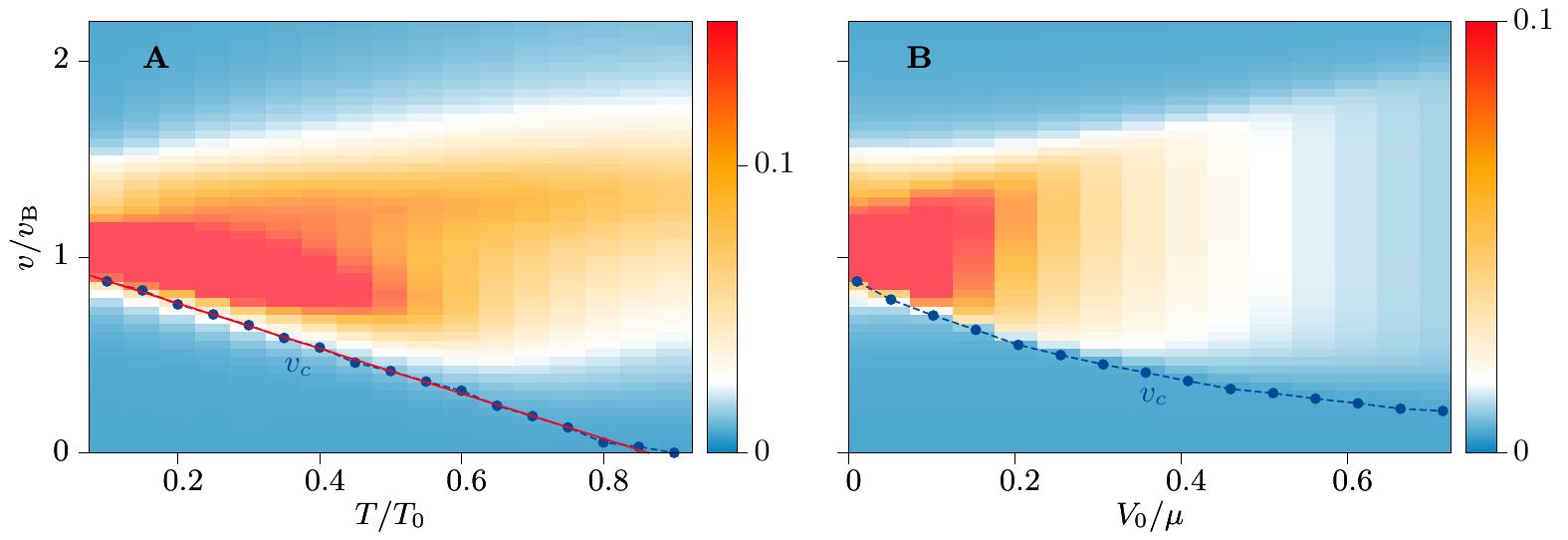}
\caption{\textbf{Influence of temperature and lattice strength on the critical velocity.}  
(\bA) Simulated heating rate $R(v)$ as a function of $T/T_0$.  
The values of the critical velocity $v_c$ are shown as circles connected with a dashed line. 
The linear fit (red continuous line) is employed to determine the critical temperature $T_c/T_0=0.86$, for which $v_c$ is zero.  
(\bB) $R(v)$ as a function of $V_0/\mu$ for $T/T_0=0.1$.  $\mu$ is the mean-field energy.  
The system parameters are  $n=1.2\, \mum^{-2}$ and $\gt=1.6$.  }
\label{FigS:v0_map}
\end{figure}

As we show in the main text, the critical velocity is smaller for high temperatures. 
In this section we analyze this thermal reduction systematically by varying the temperature $T/T_0$ in the range $0.1 -0.9$, 
where $T_0$ is the estimate of the critical temperature. 
We simulate the heating rate $R(v)$ for $n=1.2\, \mum^{-2}$, $\gt=1.6$, and  $k_0/k_c= 0.4$. 
$k_c = \sqrt{2}/\xi$ is determined by the healing length $\xi$.
In Fig. \ref{FigS:v0_map}A we show $R(v)$ as a function of $T/T_0$ for the lattice strength $V_0/\mu= 0.01$.
With increasing temperature, the broadening of the heating rate increases and the onset of rapid increase occurs at a lower velocity. 
At low temperatures the heating rate primarily shows one maximum close to the Bogoliubov velocity  $v_\mB=7.3\, \mmms$. 
At high temperatures the heating rate shows two maxima corresponding to the two sound modes.
We fit the heating rate below $v_\mB$ to the function $f(v)= A_0 \mmax[0, v^2 -v_c^2]$, with $A_0$ and $v_c$ as fitting parameters.
We show the determined values of $v_c$ in Fig. \ref{FigS:v0_map}A. 
The critical velocity decreases linearly with the temperature. 
We fit this temperature dependence of $v_c$ with a linear function to determine the critical velocity at zero temperature $v_c(0)= 7.3\, \mmms$, which is in excellent agreement with the Bogoliubov velocity $v_\mB$. 
From the fit, we also extrapolate the temperature $T/T_0= 0.86$, for which $v_c$ is zero. 
We denote this temperature as the critical temperature. 
In Fig. \ref{FigS:v0_map}B we show $R(v)$ as a function of the lattice strength $V_0/\mu$ for $T/T_0=0.1$.  
$\mu$ is the mean-field energy.  
The lattice strength introduces an additional broadening of the heating rate. 
This broadening is higher for higher $V_0/\mu$. 
We determine the value of the critical velocity $v_c$, which is shown in Fig. \ref{FigS:v0_map}B. 
The critical velocity decreases in a non-linear fashion as a function of  $V_0/\mu$.

 \section{Determining the sound velocity from the propagation of a density wave}
\begin{figure}[]
\includegraphics[width=1.0\linewidth]{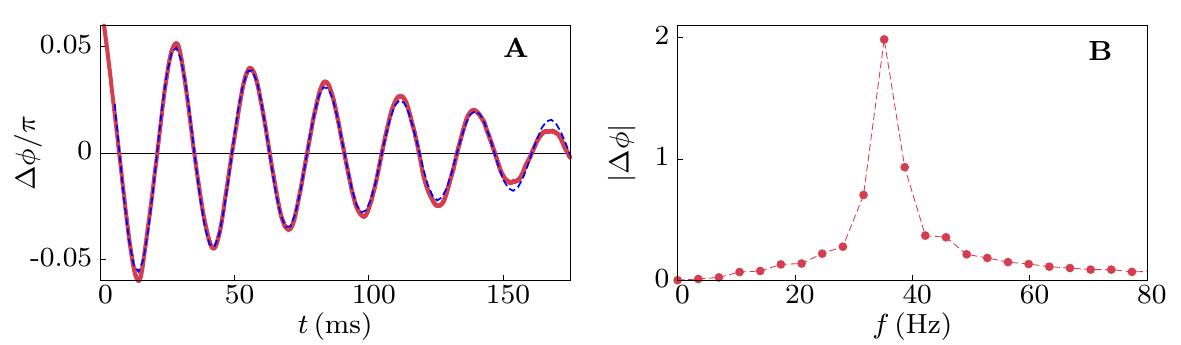}
\caption{\textbf{Sound propagation.}  
(\bA) Time evolution of the phase difference between the two subsystems, $\Delta \phi (t)$, for $n=1.2\, \mum^{-2}$, $T/T_0=0.1$, and $\gt=1.6$. 
The dashed line is the fit with a damped sinusoidal function, see text.  
(\bB) Spectrum of $\Delta \phi (t)$ shows a peak at the sound frequency.      }
\label{FigS:phase_time}
\end{figure}

To determine the sound velocity we excite a density wave following the method of Refs. \cite{Luick2020, SinghJJ}. 
We imprint a phase difference on one-half of  the system along $x$ direction. This sudden imprint of the phase results in an oscillation of the phase difference between the two subsystems, $\Delta \phi (t)$,  as shown in Fig. \ref{FigS:phase_time}A. 
We fit $\Delta \phi (t)$ with a damped sinusoidal function $f(t)= A_0 \exp(- \Gamma t) \sin(2\pi f + \phi_0)$ to determine the amplitude $A_0$, the damping rate $\Gamma$, the frequency $f$, and the phase shift $\phi_0$. 
From the fit to the results in Fig. \ref{FigS:phase_time}A we obtain $f=35.8\, \mHz$ and $\Gamma/(2\pi) =1.29 \, \mHz$. 
The sound velocity is determined by $v_s= 2fL_x$, where  $L_x$ is the system length in the $x$ direction. 
This results in $v_s= 7.16\, \mmms$ and the damping velocity $v_\Gamma=0.26  \, \mmms$, which are $v_s/v_\mB=0.98$ and  $v_\Gamma/v_\mB=0.03$, respectively. The reduction of the sound velocity is small for $T/T_0=0.1$, as expected. 
In Fig. \ref{FigS:phase_time}B we show the spectrum of $\Delta \phi (t)$, which yields the peak at the sound frequency and is the same as the one  
determined from the time evolution in Fig. \ref{FigS:phase_time}A.

 \section{Determining the superfluid density}
\begin{figure}[]
\includegraphics[width=0.5\linewidth]{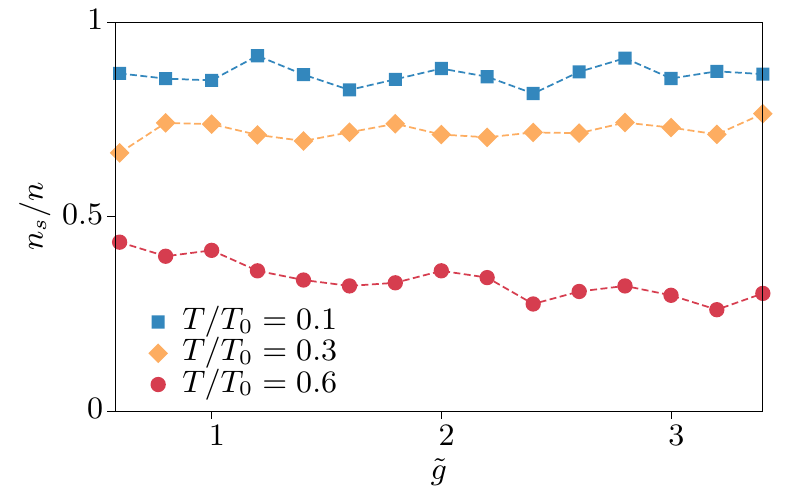}
\caption{\textbf{Superfluid density.}  
Numerical superfluid fraction $n_s/n$ as a function of the interaction strength $\gt$ for $n=1.2\, \mum^{-2}$ and $T/T_0=0.1$ (squares), $0.3$ (diamonds), and $0.6$ (circles).    }
\label{FigS:sf_gt}
\end{figure}

We determine the superfluid density by calculating the current-current correlations in momentum space, see also \cite{SinghSS, sf_2019}. 
The current density $\bj(\br)$ is defined as 
\begin{align}
\bj(\br) = \frac{\hbar}{2im} [  \psi^\ast (\br) \nabla \psi (\br)  - \psi (\br) \nabla \psi^\ast (\br)  ].
\end{align}
We choose the gradient along $x/y$ directions and calculate the Fourier transform of the current density $(j_\bk)_{x/y}$ in the $x$ and $y$ directions. This allows us to determine $\langle (j_{\bk}^\ast)_x  (j_{\bk})_y  \rangle $, which in the limit of $k \rightarrow 0$ are approximated by 
\begin{align}\label{eq:corr}
\langle (j_{\bk}^\ast)_l  (j_{\bk})_m  \rangle = \frac{k_\mB T}{m} A \Bigl( n_s \frac{k_l k_m}{k^2} + n_n \delta_{lm}  \Bigr).
\end{align}
$n_s$ and $n_n$ are the superfluid and the normal fluid density, respectively. 
$A$ is the system area. By taking a cut along the line $k_x= k_y = k/ \sqrt{2}$ and by estimating the $k=0$ value using a linear fit, we determine the superfluid density based on Eq. \ref{eq:corr}.
In Fig. \ref{FigS:sf_gt} we show the interaction dependence of the numerically determined superfluid density. 
While the superfluid density doesn't show a qualitative dependence on the interaction strength for low and intermediate temperatures, 
it decreases with increasing interaction strength for high temperature.

\end{document}